\documentclass[aps,prb,showpacs,twocolumn,floats,epsfig,pdflatex]{revtex4}
\usepackage{epsfig}
\usepackage{amsmath}
\usepackage{amsfonts}
\usepackage{graphicx}
\usepackage{amssymb}
\usepackage{amsbsy}

\newcommand{\rr}{\mathbf{r}}

\begin{document}

\title {A projection operator approach to the Bose-Hubbard model}

\author{Anirban Dutta $^{(1)}$, C. Trefzger $^{(2)}$, and K. Sengupta $^{(1)}$}

\affiliation{$^{(1)}$Theoretical Physics Department, Indian
Association for the Cultivation of Science, Jadavpur,
Kolkata-700032, India.\\
$^{(2)}$ ICFO - Institut de Ciencies Fotoniques, Mediterranean
Technology Park, 08860 Castelldefels (Barcelona), Spain.}

\date{\today}

\begin{abstract}

We develop a projection operator formalism for studying both the
zero temperature equilibrium phase diagram and the non-equilibrium
dynamics of the Bose-Hubbard model. Our work, which constitutes an
extension of Phys. Rev. Lett. {\bf 106}, 095702 (2011), shows that
the method provides an accurate description of the equilibrium zero
temperature phase diagram of the Bose-Hubbard model for several
lattices in two- and three-dimensions (2D and 3D). We show that the
accuracy of this method increases with the coordination number $z_0$
of the lattice and reaches to within $0.5\%$ of quantum Monte Carlo
data for lattices with $z_0=6$. We compute the excitation spectra of
the bosons using this method in the Mott and the superfluid phases
and compare our results with mean-field theory. We also show that
the same method may be used to analyze the non-equilibrium dynamics
of the model both in the Mott phase and near the
superfluid-insulator quantum critical point where the hopping
amplitude $J$ and the on-site interaction $U$ satisfy $z_0J/U \ll
1$. In particular, we study the non-equilibrium dynamics of the
model both subsequent to a sudden quench of the hopping amplitude
$J$ and during a ramp from $J_i$ to $J_f$ characterized by a ramp
time $\tau$ and exponent $\alpha$: $J(t)=J_i +(J_f-J_i)
(t/\tau)^{\alpha}$. We compute the wavefunction overlap $F$, the
residual energy $Q$, the superfluid order parameter $\Delta(t)$, the
equal-time order parameter correlation function $C(t)$, and the
defect formation probability $P$ for the above-mentioned protocols
and provide a comparison of our results to their mean-field
counterparts. We find that $Q$, $F$, and $P$ do not exhibit the
expected universal scaling. We explain this absence of universality
and show that our results for linear ramps compare well with the
recent experimental observations.

\end{abstract}


\maketitle

\section{Introduction}

Ultracold bosonic atoms in optical lattices provide us with an
unique setup to study properties of bosons near a Mott
insulator-superfluid (MI-SF) quantum critical
point\cite{bloch1,exp1}. A careful analysis of such experimental
bosonic systems in optical lattices show that their low-energy
properties are well described by the Bose-Hubbard model
\cite{zoller1}, which has already been theoretically studied using
both analytical \cite{fisher1,tvr1,jim1} and numerical
\cite{trivedi1} techniques. The presence of such an experimental
test bed has led to a plethora of new theoretical studies on the
model \cite{dupuis1,dupuis2,mfd1,qmc3d1,kr1}. Many of the earlier
analytical studies have concentrated on obtaining the phase diagram
of the model by using mean-field theory \cite{fisher1,tvr1},
excitation energy computation \cite{jim1}, and strong-coupling
expansion for the boson Green function \cite{kr1}. The results
obtained by these methods have been compared to extensive quantum
Monte Carlo (QMC) data \cite{trivedi1,qmc3d1}. Out of these methods,
the strong-coupling expansion \cite{kr1} (excitation energy
computation \cite{jim1}) and the NPRG approach \cite{dupuis2}
provide the closest match to QMC data in 2D (3D).

Recently, it has been realized that such ultracold bosonic systems
also allow us easy access to the non-equilibrium dynamics of its
constituent atoms near the MI-SF quantum critical point. The
theoretical study of such quantum dynamics on various models has
seen great progress in recent years \cite{rev1}. Most of these works
have either restricted themselves to the physics of integrable
and/or one-dimensional (1D) models or concentrated on generic
scaling behavior of physical observable for sudden or slow dynamics
through a quantum critical point \cite{rev1,ap1,ks1,ks2,ds1}.
However quantum dynamics of specific experimentally realizable
non-integrable models in higher spatial dimensions and strong
coupling regime has not been studied extensively mainly due to the
difficulty in handling quantum dynamics of plethora of states in the
system's Hilbert space. The Bose-Hubbard model with on-site
interaction strength $U$ and nearest neighbor hopping amplitude $J$,
which provides an accurate description for ultracold bosons in an
optical lattice, constitutes an example of such models. Most of the
studies on dynamics of this model have concentrated on $d=1$
\cite{anatoly1}, weak coupling regime \cite{gppapers1}, and
mean-field order parameter dynamics following a sudden ramp in the
strong coupling regime \cite{ehud1,ken1,others1}. Recent experiments
\cite{exp1} clearly necessitate computation of dynamical evolution
of several other quantities in higher dimensional Bose-Hubbard model
in the strong-coupling regime ($U \gg J$) beyond the mean-field
theory and for arbitrary ramp time $\tau$. However, none of the
works mentioned above presents an analysis of the non-equilibrium
dynamics of the model beyond mean-field theory.

More recently, the authors of Ref.\ \onlinecite{ct1} have developed
a theoretical formalism which enables one to analyze the dynamics of
the Bose-Hubbard model beyond mean-field theory near the MI-SF
critical point \cite{ct1}. The method uses a projection operator
technique which enables us to account for the quantum fluctuations
over the mean-field theory perturbatively in $J_f/U$($J(t)/U$) and
therefore yields accurate results as long $J_f/U (J(t)/U) \ll 1$ for
sudden(ramp) dynamics. This allows one to treat sudden and slow
ramps at equal footing near the MI-SF quantum critical point. As
shown in Ref.\ \onlinecite{ct1}, the projection operator method
yields an accurate phase diagram and also provides an estimate of
dynamically generated defect density which shows a qualitatively
reasonable match with recent experimental results \cite{exp1}. In
the present work, we extend these results in several ways. First, we
present a generic phase diagram of the Bose-Hubbard model as a
function of the lattice coordination number $z_0$ and compare these
results to the available QMC data for several one-, two-, and
three-dimensional lattices. Our comparison demonstrates that the
accuracy of the projection operator technique increases with $z_0$
reaching about $0.5\%$ of the QMC data for lattices with $z_0=6$.
Second, we compute the excitation spectrum using our approach and
show that it yields the gapless Bogoliubov and gapped amplitude
modes in the SF phase and the gapped particle and hole excitation
modes in the MI phase. Third, we study the dynamics of the model for
non-linear ramp of the hopping parameter $J$ from $J_i$ to $J_f$
characterized by a ramp time $\tau$ and exponent $\alpha$: $J(t)=J_i
+(J_f-J_i) (t/\tau)^{\alpha}$. We compute the fidelity
susceptibility $F$, nearest-neighbor correlation between the bosons
$B$, the defect formation probability $P$, and the residual energy
$Q$ of the system following such a protocol and show that our result
reproduce those of Ref.\ \onlinecite{ct1} for $\alpha=1$ as a
special case. We also find the value of the optimal $\alpha$ which
leads to minimal defect production for fast quenches (small $\tau$).
Finally, we also compute the order parameter $\Delta(t)$, the
order-parameter correlation function $C(t)$, the wavefunction
overlap $F$, and the residual energy $Q$ subsequent to a sudden
quench, discuss their properties, and provide explicit analytical
expressions for $\Delta(t)$ and $Q$. We also provide a detailed
comparison of the behavior of $\Delta(t)$ with that obtained from
Gutzwiller mean-field theory.

The plan of the rest of the work is the following. In Sec.\
\ref{eqsec}, we develop the projection operator formalism and apply
it to obtain the equilibrium phase diagram of the Bose-Hubbard model
for arbitrary $z_0$ and compute its excitation spectrum. This is
followed by Sec.\ \ref{dynsec}, where we discuss the dynamics of the
model both for sudden quench and non-adiabatic ramp of the hopping
amplitude $J$. Finally we discuss our results and conclude in Sec.\
\ref{dissec}.

\section{Formalism and equilibrium phase diagram}
\label{eqsec}

In this section, we provide a detailed exposition of the projection
operator formalism. In Sec.\ \ref{phaseboundary}, we compute the
MI-SF phase boundary using this formalism for various lattices while
in Sec.\ \ref{exspec}, we compute the low-energy excitation spectra
of the MI and the SF phases.

\subsection{Phase boundary}
\label{phaseboundary}

The Hamiltonian of the Bose-Hubbard model is
\begin{eqnarray}
{\mathcal H} &=& T+H_0, \quad  T =  \sum_{\langle {\bf r}{\bf
r'}\rangle} -J b_{{\bf
r}}^{\dagger} b_{{\bf r'}} \nonumber\\
H_0 &=& \sum_{{\bf r}} [-\mu {\hat n}_{{\bf r}} + \frac{U}{2} {\hat
n}_{{\bf r}}({\hat n}_{{\bf r}}-1) ] \label{ham1}
\end{eqnarray}
where $b_{\bf r}$ (${\hat n}_{\bf r}$) is the boson annihilation
(number) operator living on site ${\bf r}$ of a $d$-dimensional
lattice with coordination number $z_0=\sum_{\langle \rr'
\rangle_\rr} 1$, and the chemical potential $\mu$ fixes the total
number of particles. The exact solution of ${\mathcal H}$ is
difficult even numerically due to the infinite dimensionality of the
Hilbert space. A typical practice is to use the Gutzwiller ansatz
$|\psi \rangle = \prod_{{\bf r}} \sum_n c_{n}^{({\bf r})}
|n\rangle$, and solve for $c_n^{({\bf r})}$ keeping a finite number
of states $n$ around the Mott occupation number $n=\bar{n}$. This
yields the standard mean-field results with $c_{n}^{({\bf r})}=c_n$
for homogeneous phases of the model \cite{mfd1}.

\begin{figure}
\rotatebox{0}{\includegraphics*[width=\linewidth]{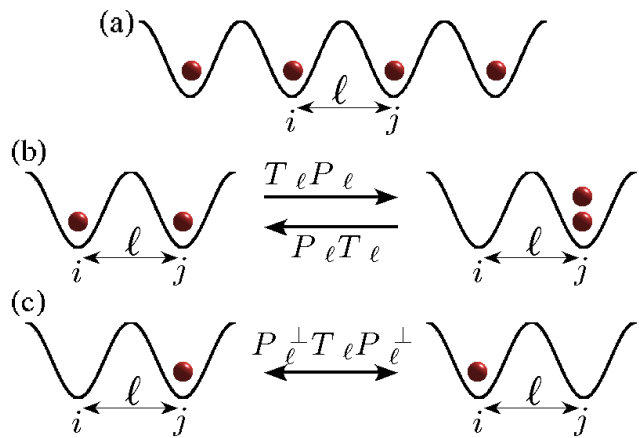}}
\caption{(Color online) Schematic representation of the Mott state
with ${\bar n}=1$. (b) Typical hopping process mediated via
$T_{\ell}^1$. (c) Hopping process mediated via $T_{\ell}^0$. Notice
that the states in (c) become part of the low-energy manifold near
the critical point, while those in the right side of (b) do not and
are always at an energy U above the Mott state.} \label{figzero}
\end{figure}

To build in fluctuations over such mean-field theory, we use a
projection operator technique \cite{issac1}. The key idea behind
this approach is to introduce a projection operator
\begin{eqnarray}
P_{\ell}= |\bar{n}\rangle \langle \bar{n}|_{\bf r} \times
|\bar{n}\rangle \langle \bar{n}|_{\bf r'} \label{projop1}
\end{eqnarray}
which lives on the link $\ell$ between the two neighboring sites
${\bf r}$ and ${\bf r'}$ of the lattice. The hopping term $T$ can
then be formally written as
\begin{eqnarray}
T &=& \sum_{\ell} T_{\ell} = \sum_{\ell} (T_\ell^0 + T_\ell^{1}) \nonumber\\
T_{\ell}^0 &=& P_{\ell}^{\perp} T_{\ell} P_{\ell}^{\perp}, \quad
T_{\ell} ^1 = (P_{\ell} T_{\ell} + T_{\ell} P_{\ell}),
\label{hopdiv}
\end{eqnarray}
where $P_{\ell}^{\perp}=(1-P_{\ell})$. The advantage of the
decomposition given by Eq.\ \ref{hopdiv} is that it distinguishes
between low- and high-energy tunneling processes as shown
schematically in Fig.\ \ref{figzero} for ${\bar n}=1$. The existence
of a low-energy subspace for the model becomes more evident by
rewriting the Bose-Hubbard Hamiltonian in the more convenient way
$\mathcal{H} = \mathcal{H}_0 + \mathcal{H}_1$, where
\begin{eqnarray}
\mathcal{H}_0 &=& H_0 + \sum_{\ell}T_\ell^{0}, \quad \mathcal{H}_1 =
\sum_{\ell}T_\ell^{1}. \label{hamnew}
\end{eqnarray}
We can then define the low-energy subspace to be a set of states
which are separated from the ground state of ${\mathcal H_0}$  by
energies $O(J)$. These set of states can not be connected to each
other by $\mathcal{H}_1$. For any two members, $|n_1\rangle$ and
$|n_2 \rangle$ of this set, one has $\langle n_1 | {\mathcal
H_1}|n_2\rangle =0$. In other words, ${\mathcal H_1}$ acting on any
state $|n_1\rangle$ in these low-energy subspace yields a state
$|n'_1\rangle$ which is necessarily separated from ground state of
${\mathcal H_0}$ by an energy ${\rm O}(U)$. Note that the states
which are member of the low-energy subspace depend on the value of
$J/U$. For example, the states schematically represented in panel
(c) of Fig.\ \ref{figzero} become members of the low-energy subspace
near the MI-SF quantum critical point where $J \simeq J_c$; however,
these states do not belong to the low-energy subspace for $J=0$.

In what follows, we shall use the projection operator technique to
systematically chart out the effective low-energy Hamiltonian by
eliminating ${\mathcal H_1}$ from ${\mathcal H}$ to ${\rm O}(J)$.
The canonical transformation operator $S$ which achieves this can be
written as
\begin{eqnarray}
S \equiv S[J]= \sum_{\ell} i[P_{\ell},T_{\ell}]/U. \label{cantrans}
\end{eqnarray}
It can be easily checked that $[iS,H_0]=-\sum_{\ell} T_{\ell}^1$ so
that the transformation eliminates $T_{\ell}^1[J]$ up to first order
in $z_0J/U$. A standard expansion in $z_0J/U$ then leads to the
effective Hamiltonian $H^{\ast}=\exp(iS) {\mathcal H} \exp(-iS)$ to
${\rm O}(z_0^2J^2/U)$
\begin{eqnarray}
H^{\ast} &=& H_0 + \sum_{\ell} P_\ell^\perp T_\ell P_\ell^\perp -
\frac{1}{U} \sum_{\ell} \Big[P_\ell T_{\ell}^2 +
 T_{\ell}^2 P_\ell \nonumber\\
&-& P_{\ell} T_{\ell}^2 P_{\ell} - T_{\ell} P_{\ell} T_\ell \Big] -
\frac{1}{U} \sum_{\langle \ell \ell^\prime \rangle} \Big[ P_{\ell}
T_\ell T_{\ell^\prime} - T_\ell P_\ell T_{\ell^\prime} \nonumber\\
&+& \frac{1}{2}\Big ( T_\ell P_{\ell} P_{\ell^\prime}
T_{\ell^\prime} - P_{\ell} T_\ell T_{\ell^\prime} P_{\ell^\prime}
\Big)  + {\rm h.c.} \Big] \label{ham2}
\end{eqnarray}
Note that the second order terms in $H^{\ast}$ involves effective
hopping processes between adjacent links leading to spatial
correlation between next-nearest neighbor sites; higher order terms
in $z_0J/U$ systematically build such correlations between further
neighbors. In this work, we restrict ourselves to ${\rm
O}[(z_0J/U)^2]$.

Using $H^{\ast}$ one can now compute the variational ground state
energy
\begin{eqnarray}
E &=& \langle \psi|{\mathcal H} | \psi \rangle = \langle
\psi'|H^{\ast}| \psi'\rangle + {\rm O}(z_0^3 J^3/U^2),
\label{encomp1}
\end{eqnarray}
where $|\psi'\rangle = \exp(iS) |\psi\rangle$, and we use a
Gutzwiller ansatz $|\psi'\rangle = \prod_{\bf r} \sum_n f_n^{({\bf
r})} |n\rangle$,
%\begin{eqnarray}
%|\psi'\rangle = \prod_{\bf r} \sum_n f_n^{({\bf r})} |n\rangle,
%\label{guzan1}
%\end{eqnarray}
for the variational wavefunction $|\psi\rangle$ in the Mott limit
($S,J=0$), where $|\psi\rangle=|\psi'\rangle$. Note that
$|\psi\rangle$ is not of the Gutzwiller form; it incorporates
spatial correlation via $\exp(iS)$ factor. To obtain the variational
energy $E$ in terms of the coefficients, we define the fields
\begin{eqnarray}
\varphi_\rr &=& \langle{\psi^\prime}| b_\rr |{\psi^\prime}\rangle =
\sum_n \varphi_{\rr n} = \sum_n \sqrt{n+1} f_\mathrm{n}^{*({\bf r})}
f_\mathrm{n+1}^{({\bf r})} \nonumber\\
\Phi_\rr &=& \langle{\psi^\prime}| b_\rr^2 |{\psi^\prime}\rangle =
\sum_n \Phi_{\rr n} \nonumber\\
&=& \sum_n \sqrt{(n+1)(n+2)} f_\mathrm{n}^{*({\bf r})}
f_\mathrm{n+2}^{({\bf r})} \label{phieq}
\end{eqnarray}
%\begin{eqnarray}
%\varphi_{\bf r}[\Phi_{\bf r}] &=& \langle{\psi^\prime}| b_{\bf r}|
%{\psi^\prime}\rangle [\langle{\psi^\prime} b_{\bf r}^2
%\rangle{\psi^\prime}] = \sum_n \varphi_{{\bf r}n}[\Phi_{{\bf r} n}]
%\nonumber\\
%\phi_{{\bf r} n} &=& \sqrt{n+1} f_\mathrm{n}^{*({\bf r})}
%f_\mathrm{n+1}^{({\bf r})} \nonumber\\
%\Phi_{{\bf r} n} &=&  \sqrt{(n+1)(n+2)} f_\mathrm{n}^{*({\bf r})}
%f_\mathrm{n+2}^{({\bf r})} \label{phieq}
%\end{eqnarray}
Using the expressions of  $\varphi_\rr$ and $\Phi_\rr$ in Eq.\
\ref{phieq}, one obtains, after some algebra, the expression for the
variational energy $E \equiv E[\{f_n\};J]$ to be
\begin{widetext}
\begin{eqnarray}
\label{en1} E &=& \sum_{{\bf r},n} [-\mu n + U n(n-1)/2 ]|f_n^{({\bf
r})}|^2  - J\sum_{\langle {\bf r} {\bf r'} \rangle}\Big \{
\varphi_{{\bf r}}^* \varphi_{{\bf r'}} - 2\Re{\varphi_{{\bf
r},\bar{n}-1}^* \varphi_{{\bf r'}\bar{n}}} + J \bar{n}(\bar{n}+1)/U
\Big[
|f_{\bar{n}}^{({\bf r})}|^2 |f_{\bar{n}}^{({\bf r'})}|^2  \nonumber\\
&& - |f_{\bar{n}+1}^{({\bf r})}|^2 |f_{\bar{n}-1}^{({\bf r'})}|^2 -
f_{\bar{n}+1}^{*({\bf r})} f_{\bar{n}-1}^{({\bf r})}
f_{\bar{n}-1}^{*({\bf r'})} f_{\bar{n}+1}^{({\bf r'})}  \Big] +2J/U
\Re{\Phi_{{\bf r},\bar{n}-2}^* \Phi_{{\bf r'}\bar{n}}}\Big\}-
J^2/U\sum_{\langle {\bf r} {\bf r'} {\bf r''} \rangle} \Big\{ 2\Re
\Big[ \varphi_{{\bf r},\bar{n}-1}^*(\bar{n}+1)
|f_{\bar{n}}^{({\bf r'})}|^2\nonumber \\
&& + \varphi_{{\bf r}\bar{n}} \Phi_{{\bf r'},\bar{n}-2}^* -
\varphi_{{\bf r}\bar{n}}^* \bar{n} |f_{\bar{n}-1}^{({\bf r'})}|^2 -
\varphi_{{\bf r},\bar{n}-1} \Phi_{{\bf r'},\bar{n}-1}^* \Big]
\varphi_{{\bf r''}} + 2\Re \Big[ \varphi_{{\bf r},\bar{n}-1}^*
\Phi_{{\bf r'}\bar{n}} + \varphi_{{\bf r}\bar{n}}
\bar{n}|f_{\bar{n}}^{({\bf r'})}|^2
- \varphi_{{\bf r}\bar{n}}^* \Phi_{{\bf r'},\bar{n}-1}\nonumber \\
&&  - \varphi_{{\bf r},\bar{n}-1}(\bar{n}+1) |f_{\bar{n}+1}^{({\bf
r'})}|^2  \Big] \varphi_{{\bf r''}}^* + \varphi_{{\bf r}\bar{n}}^*
\bar{n} \Big[ |f_{\bar{n}-1}^{({\bf r'})}|^2 - |f_{\bar{n}}^{({\bf
r'})}|^2\Big] \varphi_{{\bf r''}\bar{n}} + \varphi_{{\bf
r},\bar{n}-1} (\bar{n}+1) \Big[|f_{\bar{n}+1}^{({\bf r'})}|^2 -
|f_{\bar{n}}^{({\bf r'})}|^2 \Big] \varphi_{{\bf r''},\bar{n}-1}^*
\nonumber \\
&& + 2 \Re \varphi_{{\bf r}\bar{n}}^* \Phi_{{\bf r'},\bar{n}-1}
\varphi_{{\bf r''},\bar{n}-1}^* \Big\}, \label{eneq1}
\end{eqnarray}
\end{widetext}
Note that the first three terms in the first line of Eq.\
\ref{eneq1} represent the mean-field energy functional, while the
rest are corrections due to quantum fluctuations. Thus the
projection operator method involves a systematic way of
incorporating quantum fluctuations over the mean-field theory and we
expect the results from this method to be accurate for larger $z_0$
where mean-field theory provides an accurate starting point.

\begin{figure}
\rotatebox{0}{\includegraphics*[width=\linewidth]{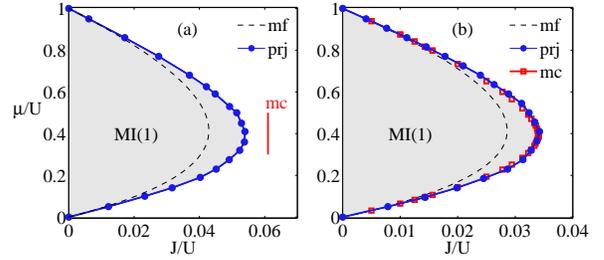}}
\caption{(Color online) Phase diagram of the Bose-Hubbard model in
2D (a) and 3D (b). The blue dots and blue solid lines (black dashed
line) indicate the phase diagram obtained by the projection operator
(mean-field) method. The red squares indicate QMC data.}
\label{fig1}
\end{figure}

The MI-SF phase diagram can be obtained by minimizing $E[\{f_n\};J]$
with respect to $\{f_n\}$ or by solving $ i\hbar
\partial_t |\psi' \rangle = H^{\ast}[J] |\psi'\rangle$ in imaginary
time \cite{comment1}. In this work, we are going to use the former
technique and restrict ourselves to ${\bar n}=1$. Such phase
diagrams  for 2D and 3D square lattice are shown in Fig.\
\ref{fig1}(a) and Fig.\ \ref{fig1}(b) respectively. We note that the
match with QMC data \cite{qmc3d1} is nearly perfect for 3D
(Fig.\ref{fig1}(b)) where mean-field theory provides an accurate
starting point. While in 3D the accuracy with QMC at the tip of the
Mott lobe is $\sim 0.5 \%$, in 2D we find $J_c/U=0.055$ compared to
the QMC value $0.061$ \cite{trivedi1} (red line in Fig.\
\ref{fig1}(a)). Here the match with QMC is not as accurate as in 3D;
however it compares favorably to other analytical methods
\cite{jim1}.

\begin{figure}
\rotatebox{0}{\includegraphics*[width=\linewidth]{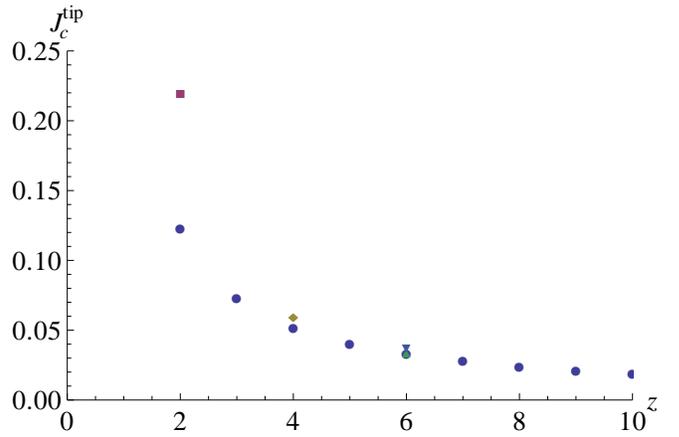}}
\caption{(Color online) The plot of $J_c^{\rm tip}$ as a function of
$z_0$ as shown by blue circles. The red square, brown hexagon, green
triangle, and the blue inverted triangle represents QMC data for 1D
Bose Hubbard model ($z_0=2$), 2D square lattice ($z_0=4$), 3D cubic
lattice ($z_0=6$), and 2D triangular lattice ($z_0=6$)respectively.}
\label{fig1a}
\end{figure}

To provide an accurate comparison of our method with other lattices,
we note that the nature of the lattice affects $J_c$ only through
$z_0$. We also note from Fig.\ \ref{fig1}(b) that the deviation of
$J_c$ from QMC value is maximal at the tip of the Mott lobe. Thus to
elucidate the $z_0$ and hence the lattice dependence of $J_c$
computed by the present method, we plot $J_c^{\rm tip}$ ({\it i.e.},
the value of $J_c$ at the tip of the Mott lobe)  as a function of
$z_0$ in Fig.\ \ref{fig1a}. The comparison of corresponding QMC data
for various lattices show that the method indeed becomes more
accurate with increasing $z_0$.

Before ending this section, we would like to note that the inclusion
of fluctuation in our method becomes apparent on computing the
expectation $\langle T_{\ell} \rangle = -J \langle \psi|b_{\bf
r}^{\dagger} b_{\bf r'} + {\rm h.c}|\psi\rangle$ in the MI phase.
The mean-field theory provide a zero result for $\langle T_{\ell}
\rangle$ , while the projection operator method yields
\begin{eqnarray}
\langle \psi|{T_\ell} | \psi \rangle &=&\langle
\psi^{\prime}|\exp(iS[J]) {T_\ell} \exp(-iS[J]) | \psi^{\prime}
\rangle\nonumber\\
&=&\langle \psi^{\prime}|{T_\ell} | \psi^{\prime} \rangle
-\frac{1}{U}  \langle \psi'|\left[P_{\ell}T_{\ell}^2+T_{\ell}^2
P_{\ell}-2T_\ell P_\ell T_\ell \right] \nonumber\\
&&+ \sum_{\langle{\ell}^{\prime}\rangle} \langle
\psi^{\prime}|\left[P_\ell T_\ell T_{\ell^{\prime}}-T_\ell P_\ell
T_\ell^{\prime}+\rm h.c.\right]|\psi'\rangle \label{kecal}
\end{eqnarray}
where we have kept terms up to ${\rm O}(J^2/U^2)$ and $\langle
\ell'\rangle$ denotes nearest neighbor links to $\ell$. In the MI
phase, the first term of Eq.\ \ref{kecal}, which is also the
mean-field result, vanishes, while the second fluctuation
contribution from the remaining terms yields $ \langle T_l \rangle =
2 J^2 {\bar n} ({\bar  n}+1)/U$ in the homogeneous limit. We note
that this agrees with fluctuation calculations of Ref.\
\onlinecite{jim1}.

\subsection{Excitation Spectrum}
\label{exspec}

To obtain the low-energy excitations, we consider a variational form
for $|\psi'\rangle$ which corresponds to perturbation over the
ground state value. This is given by
\begin{eqnarray}
|\psi'\rangle &=& \prod_{\bf r} \sum_n f_{n}^{(\rr)}(t) \nonumber\\
f_{n}^{(\rr)}(t) &=& [f_{n}^{(0)}+ \delta f_{n}^{(\rr)}(t)] e^{i
\omega_0 t} \label{exc1}
\end{eqnarray}
where $\delta f_n^{(\bf r)}(t)$ represents small perturbation over
the ground state value $f_n^{(0)}$ and can be expressed in momentum
space as
\begin{eqnarray}
\delta f_{\rm n}^{(\rr)}(t) &=& u^n_{{\bf k}} e^{i({\bf k} \cdot
{\bf r}-\omega t)} + v^n_{{\bf k}}  e^{-i({\bf k} \cdot {\bf r} -
\omega t)}. \label{exc2}
\end{eqnarray}
Substituting Eqs.\ \ref{exc1} and \ref{exc2} in the Schrodinger
equation $i \hbar \partial_t |\psi'\rangle = H^{\ast}
|\psi'\rangle$, we obtain a set of equations for $u^n_{{\bf k}}$ and
$v^n_{{\bf k}}$ which is given by
\begin{eqnarray}
\label{exc3} \hbar\omega_{\bf k} \left(
    \begin{array}{c}
       \vec{u}_{\bf k}\\
       \vec{v}_{\bf k}
    \end{array}
\right) &=& \left(
    \begin{array}{cc}
        A_{\bf k} & B_{\bf k}\\
       -B_{\bf k} & -A_{\bf k}
    \end{array}
\right) \left(
    \begin{array}{c}
       \vec{u}_{\bf k}\\
       \vec{v}_{\bf k}
    \end{array}
\right) \;.
\end{eqnarray}
Here the $\vec{u}_{\bf k}$ and $\vec{v}_{\bf k}$ are vectors with
components $u^n_{{\bf k}}$ and $v^n_{{\bf k}}$ ($n=0,1,\dots$),
respectively and $A_{\bf k}$ and $B_{\bf k}$ are square matrices
with elements $A_{\bf k}^{mn}$ and $B_{\bf k}^{nm}$. Since $0\le n
\le \infty$, in principle, Eq.\ \ref{exc3} represents an
infinite-dimensional matrix equation; however, in the strong
coupling regime where states with $n > 3$ bosons are energetically
costly, it is possible to truncate the arrays ${\vec u_{\bf k}}$ and
${\vec v_{\bf k}}$ by putting $u^n_{{\bf k}}, v^n_{{\bf k}} =0$ for
$n>3$. In this case the column vector $({\vec u_{\bf k}}, {\vec
v_{\bf k}})^T$ can be written as $(u_{\bf k}^0, u_{\bf k}^1, u_{\bf
k}^2, u_{\bf k}^3, v_{\bf k}^0, v_{\bf k}^1, v_{\bf k}^2, v_{\bf
k}^3)^T$. Thus the solution for the excitation spectrum reduces to
the solution of a $8 \times 8$ matrix for each ${\bf k}$. In what
follows we provide an analytical solution for $\omega_{\bf k}$ in
the MI phase and a numerical plot of the excitation spectrum in the
SF phase, where the algebra, for reasons mentioned below, turns out
to be too complicated to allow a straightforward analytical result.
We note here that our analysis amounts to generalization of the work
in Ref.\ \onlinecite{exct1} which provides the excitation spectrum
of the Bose-Hubbard model using mean-field theory.

\begin{figure}
\rotatebox{0}{\includegraphics*[width=\linewidth]{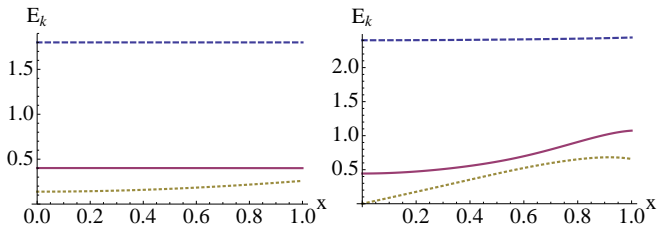}}
\caption{(Color online) Left panel: Plot of the excitation spectra
$E_{{\bf k} 1}$ (dashed blue line), $E_{{\bf k} 2}$ (solid purple
line) and $E_{{\bf k} 3}$ (dotted yellow line) as a function of $x$
in the MI phase for $\mu=0.4 U$ and $J=0.15 J_c$. Right panel:
Analogous plots for the excitation branches in the superfluid phase
showing the gapless Bogoliubov mode and the gapped amplitude modes
for $\mu=0.4 U$ and $J=1.3 J_c$. } \label{figex1}
\end{figure}

In the MI phase, $f_{n}^{(0)} = \delta_{n1}$ and $\hbar \omega_0$
can be shown to correspond to the ground state energy of the
homogeneous MI state as obtained by putting $f_1^{(0)}=1$ in Eq.\
\ref{eneq1}: $\hbar \omega_0 = -\mu -4z_0J^2/U$. Further, one finds
that in the MI phase the elements $A_{\bf k}^{nm} \sim \delta_{nm}$.
Using the expression of $\omega_0$, these diagonal elements can be
calculated to be
\begin{eqnarray}
A_{\bf k}^{00} &=&\mu - Jz_0 \left(1-2 x^2\right) -\frac{2 J^2
z_0}{U} \left[\left(1-2 x^2\right)^2
z_0-3\right]\nonumber\\
A_{\bf k}^{11}&=&-\frac{4 J^2 \left(1-2 x^2\right) z_0}{U}\nonumber\\
A_{\bf k}^{22}&=& U-2\mu-J z_0\left(1-2 x^2\right) + A_{\bf k}^{00}
\nonumber\\
A_{\bf k}^{33}&=&\frac{4 J^2 z_0}{U}+3 U-2 \mu. \label{diagel}
\end{eqnarray}
Note that in the limit $J=0$, these elements correspond to the
on-site excitation energies of the different $|n\rangle$ states. The
off-diagonal elements are given by
\begin{eqnarray}
B_{\bf k}^{06} &=&- \frac{3 \sqrt{2} J^2 z_0 \left(\left(1-2
x^2\right)^2 z_0-1\right)}{U} = B_{\bf k}^{24} \nonumber\\
B_{\bf k}^{1 5}&=&-\frac{4 J^2 \left(1-2 x^2\right) z_0}{U}.
\label{odiagel}
\end{eqnarray}
where $x= \sum_{a=1..d} \sin^2(k_a/2)/d$ and we have set the lattice
spacing to unity. Diagonalization of the matrix in Eq.\ \ref{exc3}
leads to the excitation spectra given by
\begin{widetext}
\begin{eqnarray}
E_{{\bf k} 1} &=& 3 (U-\mu ), \quad E_{{\bf k} 2} =
\frac{\sqrt{\left(4 J^2 z_0+U \mu \right)
\left[U\mu- 4 J^2 \left(4 x^2-3\right)\right]}}{U}, \nonumber\\
E_{{\bf k} 3} &=&  \frac{1}{2} \left[\left(\frac{-56 J^4 z_0^2}{U^2}
\left(\left(1-2 x^2\right)^2
   z_0-1\right)^2 + \left(J^2 z_0 \left(\left(1-2 x^2\right)^2
   z_0+8\right)+12 J \left(1-2 x^2\right) z_0 \mu +4 \mu
   ^2\right)\right.\right. \nonumber\\
&& \left.\left.\frac{-8 J^2
   z_0}{U} \left(\left(1-2 x^2\right)^2 z_0-1\right) \left(3 J \left(2
   x^2-1\right) z_0-2 \mu \right) + 2 U \left(3 J \left(2 x^2-1\right)
   z_0-2 \mu \right)+U^2\right)\right]^{1/2} \nonumber\\
&& + J z_0 \left(2 x^2-1\right)+U-2 \mu
\end{eqnarray}
\end{widetext}
Note that $E_{{\bf k}2}$ corresponds to the hole branch while
$E_{{\bf k} 3}$ corresponds to the particle branch. The energies of
these branches differ from their mean-field counterparts in Ref.\
\onlinecite{exct1} via presence of additional fluctuation
contribution which manifest then through ${\rm O}(J^2/U)$ terms. The
plots of these excitation energies as a function of $x$ is shown in
the left panel of Fig.\ \ref{figex1} and matches qualitatively with
its counterpart in Ref.\ \onlinecite{exct1}.

A similar analysis for the superfluid phase can easily be carried
out using the same algorithm described above. In this case, it turns
out that the analytical expressions of $A_{\bf k}$ ( which now has
off-diagonal terms) and $B_{\bf k}$ are prohibitively lengthy. We
therefore resort to numerical solution of Eq.\ \ref{exc3} for
several values of $x$. The result is shown in the right panel of
Fig.\ \ref{figex1}. The qualitative features of the plots are again
similar to the mean-field results of Ref.\ \onlinecite{exct1};
however quantitative values of physical quantities such as the
velocity of the Bogoliubov mode, $v_g$, differ. The difference with
the mean-field result comes, again, from the presence of ${\rm
O}(J^2/U)$ terms in the effective action and hence is small near the
critical point.

\section{Dynamics}
\label{dynsec}

We now demonstrate that the method elaborated in Sec.\ \ref{eqsec}
with minor modification allows one to address the dynamics of the
Bose-Hubbard model. To this end, we are going to assume a protocol
where the hopping amplitude $J\equiv J(t)$ changes in time from its
initial value $J_i$ to some final value $J_f$. The necessary
condition for our method to yield accurate result, as we shall
demonstrate, is $z_0J(t)/U \ll 1$ at all times.

We begin with the Schr\"odinger equation for the time dependent
Hamiltonian ${\mathcal H}[J(t)]$ which is given by
\begin{eqnarray}
i \hbar \partial_t |\psi \rangle = {\mathcal H}[J(t)] |\psi\rangle
\label{sch1}
\end{eqnarray}
The solution of this equations is difficult due to the infinite
dimensionality of the bosonic Hilbert space. However, one notes that
the contribution to the dynamics of the bosons, as long as
$z_0J(t)/U \ll 1$, comes from a limited set of states which are
members of the {\it instantaneous} low-energy subspace at any given
instant $t$. To capture the contribution of the states in this
low-energy subspace, we make a time-dependent transformation
$|\psi'\rangle = \exp(iS[J(t)]) |\psi\rangle$, which eliminates
$T_{\ell}^0[J(t)]$ up to first order from ${\mathcal H}[J(t)]$ at
each instant, and leads to the effective Hamiltonian $H^{\ast}[J(t)]
=\exp(iS[J(t)]) {\mathcal H}[J(t)] \exp(-iS[J(t)])$. This yields the
equation
\begin{eqnarray}
(i \hbar \partial_t + \partial S/\partial t)|\psi'\rangle
=H^{\ast}[J(t)]|\psi'\rangle. \label{sch2}
\end{eqnarray}
We note that the additional term $\partial S/\partial t$ takes into
account the possibility of creation of excitations during the time
evolution with a finite ramp rate $\tau^{-1}$. The above equation
yields an accurate description of the ramp with $H^{\ast}[J(t)]$
given by Eq.\ \ref{ham2} for $J(t)/U\ll 1$. Note that this does not
impose a constraint on magnitude of $\tau$; it only restricts
$J_f/U$ and $J_i/U$ to be small. Thus the method can treat both
``slow" and ``fast" ramps at equal footing. Substituting
$|\psi'\rangle = \prod_{\bf r} \sum f_n^{({\bf r})} |n\rangle$, we
obtain a set of coupled equations for the coefficients $\{f_n \}$
\begin{multline}
i\hbar \frac{\partial f_{\rm n}^{(\rr)}}{\partial t} =
\frac{i\hbar}{U} \frac{\partial J}{\partial t}
\Big[ \sqrt{n}f_{\rm n-1}^{(\rr)}\overline{\partial\alpha}_\rr
+ \sqrt{n+1}f_{\rm n+1}^{(\rr)} {\overline{\partial\beta}_\rr}^* \Big] \\
+ \chi_{\rm n}^{(\rr)} f_{\rm n}^{(\rr)} - J \Big[\sqrt{n}f_{\rm n-1}^{(\rr)}
\overline{\alpha}_\rr + \sqrt{n+1}f_{\rm n+1}^{(\rr)} {\overline{\beta}_\rr}^*\Big ]
\\ - \frac{J^2}{U} \Big[\sqrt{n(n-1)}f_{\rm n-2}^{(\rr)} \overline{\eta}_\rr
 + \sqrt{(n+1)(n+2)}f_{\rm n+2}^{(\rr)} {\overline{\xi}_\rr}^*\Big],
\label{feq1}
\end{multline}
where the fields $\chi_{\rm n}^{(\rr)}, \overline{\alpha}_\rr,
{\overline{\beta}_\rr}^*, \overline{\eta}_\rr,
{\overline{\xi}_\rr}^*, \overline{\partial\alpha}_\rr$, and
${\overline{\partial\beta}_\rr}^*$ have to be calculated
self-consistently and are explicitly given in the
Appendix~\ref{app:fields}. Notice that the last line of
Eq.~\ref{feq1} couples the time derivative of the coefficient
$f_{\rm n}^{(\rr)}$ with the coefficients $f_{\rm n\pm2}^{(\rr)}$,
the coupling being proportional to $J^2/U$. In particular this is
different than the standard mean-field equations where the time
derivative of the coefficient $f_{\rm n}^{(\rr)}$ is at most coupled
to the coefficients $f_{\rm n\pm1}^{(\rr)}$ through $J$. It is worth
noting that Eq.~\ref{feq1} conserve the total number of particles
for any $J(t)$.
%\begin{eqnarray}
%i \hbar \partial_t f_n^{({\bf r})} &=& \delta E[\{f_n(t)\};J(t)]/
%\delta f_n^{\ast ({\bf r})} + \frac{i\hbar}{U} \frac{\partial J(t)}{\partial t}\label{feq1} \\
%&\times& \sum_{\langle  {\bf r'} \rangle_{\bf r}} \sqrt{n}
%f_{n-1}^{({\bf r})}
%\Big[\delta_{n\bar{n}}\varphi_{{\bf r'}\bar{n}} - \delta_{n,\bar{n}+1}\varphi_{{\bf r'},\bar{n}-1}\Big] \nonumber \\
%&& + \sqrt{n+1} f_{n+1}^{({\bf r})} \Big[
%\delta_{n\bar{n}}\varphi_{{\bf r'},\bar{n}-1}^* -
%\delta_{n,\bar{n}-1} \varphi_{{\bf r'}\bar{n}}^* \Big]. \nonumber
%\end{eqnarray}
In what follows, we shall obtain a numerical solution of Eq.\
\ref{feq1} to address the dynamics of a translationally invariant
Bose-Hubbard model both for sudden quench and non-linear ramp of
$J(t)$.

\subsection{Sudden Quench}
\label{suudq}

In this section, we are going to address the dynamics of the bosons
after a sudden quench of the hopping amplitude from $J_i$ (Mott
phase) to $J_f$ (superfluid phase) through the tip of the Mott lobe
where the dynamical critical exponent $z=1$. Our main objective here
is to compute the time evolution of the order parameter $\Delta_{\bf
r}(t) = \langle \psi(t)|b_{\bf r}|\psi(t)\rangle$ and the
order-parameter correlation function $C_{\bf r}(t) = \langle
\psi(t)|b_{\bf r} b_{\bf r} |\psi(t)\rangle -\Delta_{\bf r}^2(t)$.
We shall also consider sudden quenches which start at the critical
point ($J_i=J_c$) and end in the superfluid phase $J_f > J_c$, and
compute the resultant residual energy $Q$ and the wavefunction
overlap $F$.

\begin{figure}
\rotatebox{0}{\includegraphics*[width=\linewidth]{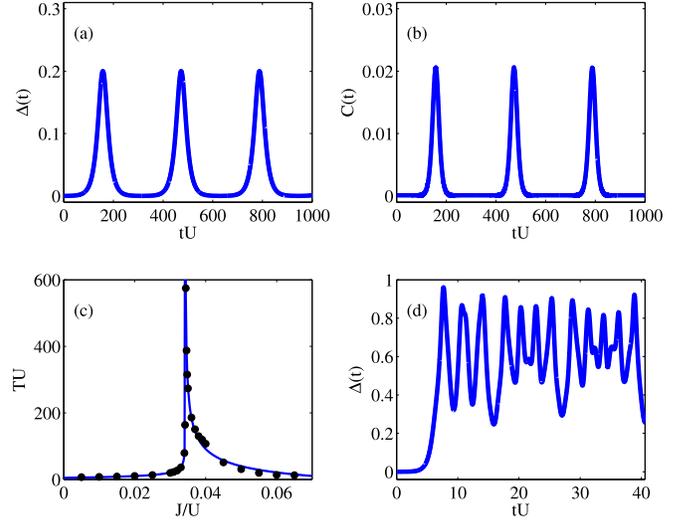}}
\caption{(Color online) Plot of $\Delta(t)$ (a), and $C(t)$ (b) as a
function of $tU$, for $J_f=1.02J_c$.  (c) The time period $T$ of the
oscillations of $\Delta(t)$. (d) Same as in (a) for $J_f=3.51J_c$.
We have set $\hbar=1$ for all plots.} \label{fig2}
\end{figure}

To this end, we begin by noting that for a sudden quench, $\partial
J/\partial t \sim \delta(t)$ and thus the first term on the right
side of Eq.\ \ref{feq1} does not contribute to the subsequent
time-evolution of the system for $t>0$.
%To solve the
%Schr\"odinger equation $i \hbar
%\partial_t  |\psi \rangle = H |\psi\rangle$, we therefore make the
%transformation $|\psi'\rangle = \exp(iS[J_f]) |\psi\rangle$ leading
%to
%\begin{eqnarray}
%i\hbar \partial_t |\psi' \rangle = H^{\ast}[J_f]
%|\psi'\rangle.\label{feq1a}
%\end{eqnarray}
%Here $|\psi'(t=0)\rangle$ is the ground state of $H^{\ast}[J_i]$ as
%determined by minimization of $E[\{f_n\};J_i]$ (Eq.\ \ref{eneq1}).
%From this, we obtain the set of coupled equations for $f_n^{({\bf
%r})}$:
%\begin{eqnarray}
%i \hbar \partial_t f_n^{({\bf r})}(t) = \delta
%E[\{f_n(t)\};J_f]/\delta f_n^{\ast ({\bf r})} \label{feq1b}
%\end{eqnarray}
%which needs to be solved numerically.
The time evolution of the order parameter $\Delta(t)$ can then be
written in terms of $\{f_n(t)\}$ by noting that $\Delta(t) = \langle
\psi'(t)|b'_{\bf r}|\psi'(t)\rangle$, where $b_{\bf r}'=
\exp(iS[J_f]) b_{\bf r} \exp(-iS[J_f])$. One can then express
$\Delta(t)$ in terms of $f_n^{({\bf r})}$ as
\begin{eqnarray}
\Delta_{\bf r}(t) &=& \varphi_{\bf r}(t) + J/U \sum_{\langle {\bf
r'} \rangle_{\bf r}} \bar{n}
\Big[|f_{\bar{n}}^{({\bf r})}|^2 - f_{\bar{n}-1}^{({\bf r})}|^2 \Big] \varphi_{{\bf r'}\bar{n}} \nonumber \\
&+& (\bar{n}+1) \Big[|f_{\bar{n}}^{({\bf r})}|^2 -
f_{\bar{n}+1}^{({\bf r})}|^2 \Big]  \varphi_{{\bf r'},\bar{n}-1}
+ \Big[\Phi_{{\bf r},\bar{n}-2} \nonumber \\
&-&  \Phi_{{\bf r},\bar{n}-1} \Big] \varphi_{{\bf r'}\bar{n}}^* +
\Big[\Phi_{{\bf r}\bar{n}} - \Phi_{{\bf r},\bar{n}-1} \Big]
\varphi_{{\bf r'},\bar{n}-1}^*. \label{odyn1}
\end{eqnarray}
Note that the first term in Eq.\ \ref{odyn1} represents the
mean-field result while the presence of the other terms indicate
contribution from the quantum fluctuations from mean-field theory.
The role of such quantum fluctuations in the evolution of
$\Delta_{\bf r}(t)$ becomes evident in computing the equal-time
order parameter correlation function $C_{\bf r}(t)$. To compute
$\Delta_{\bf r}$ and $C_{\bf r}$, we consider a spatially
homogeneous system and solve the Schr\"odinger equation (Eq.\
\ref{feq1}) for $f_n^{({\bf r})} \equiv f_n$ (as guaranteed by
translational invariance) keeping all states for $0\le n\le 5$ with
$\bar n=1$. The resultant plot of $\Delta_{\bf r}(t) \equiv
\Delta(t)$ is shown in Fig.\ \ref{fig2}(a)[(d)] for $J_i=0$ and
$J_f/J_c=1.02$($J_f/J_c=3.51$). We find that near the critical
point, $\Delta (t)$ displays oscillations with a single
characteristic frequency \cite{ehud1} while away from the critical
point ($J_f/J_c=3.51$), multiple frequencies are involved in its
dynamics. The time period $T$ (Fig.\ \ref{fig2}(c)) of these
oscillations near $J_c$ is found, as a consequence of critical
slowing down, to have a divergence $T \sim (\delta J)^{-0.35\pm
0.05}$ leading to $z\nu = 0.35\pm 0.05$ for $d=3$
\cite{rev1,comment2}. Finally, we plot $C_{\bf r}(t) \equiv C(t)$ as
a function of $t$ for $J_f=1.02J_c$ in Fig.\ \ref{fig2}(b). We find
that $|C(t)/\Delta^2(t)|$ may be as large as $0.5$ at the tip of the
peaks of $\Delta(t)$, which shows strong quantum fluctuations near
the QCP.

\begin{figure}
\rotatebox{0}{\includegraphics*[width=\linewidth]{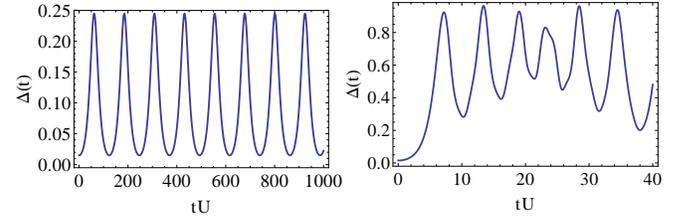}}
\caption{(Color online) Plot of $\Delta(t)$ as a function of $Ut$
($\hbar =1$) as computed using mean-field theory for
$J_f=1.02J_c^{\rm mf}$ (left panel) and $J_f=3.51 J_c^{\rm mf}$
(right panel).} \label{figdeltamf}
\end{figure}

To compare our results with the order parameter dynamics obtained
from the mean-field theory, we solve the equations of motion for the
time dependent Gutzwiller coefficients $f_n(t)$ within a single-site
mean-field theory. As shown in Ref.\ \onlinecite{ehud1,others1}, the
mean-field equation reads
\begin{eqnarray}
(i \partial_t -\epsilon_n) f_n &=& -z_0 J(t) \Big[ \Delta_{\rm
mf}(t)
\sqrt{n} f_{n-1} \nonumber\\
&& + \Delta_{\rm mf}^{\ast}(t) \sqrt{n+1} f_{n+1} \Big] \label{mft1}
\end{eqnarray}
where $\epsilon_n = -\mu n +Un (n-1)/2$ is the on-site energy of the
bosons, $\Delta_{\rm mf} (t) = \sum_n f_{n-1}^{\ast} f_n \sqrt{n}$,
and $J(t)= (J_i \theta(-t)+J_f \theta(t))$ for the sudden quench
protocol. For the mean-field theory the critical point lies at
$J_c^{\rm mf} =0.028 U$ for $d=3$. To obtain the order parameter
dynamics, we obtain the values of $f_n(t)$ numerically keeping up to
$n=5$ states and compute the order parameter $ \Delta_{\rm mf}(t)$
choosing the same sudden quench protocol used in the projection
operator approach (see Fig.\ \ref{fig2}). The behavior of
$\Delta_{\rm mf}(t)$ as a function of time is shown in the left
panel Fig.\ \ref{figdeltamf}. A comparison of our results with that
of the mean-field theory can now be made by comparing Figs.\
\ref{fig2} and \ref{figdeltamf}. We find that although the
qualitative nature of $\Delta_{\rm mf}(t)$ and $\Delta(t)$ are
similar, the periodicity of the oscillations are quite different.
Further we note that
\begin{eqnarray}
C_{\rm mf}(t)&=& \langle b_{\bf r}^2 \rangle -\Delta^2_{\rm mf}(t)
\nonumber\\
&=& \sum_n \sqrt{n(n-1)} f_{n-2}^{\ast} f_n -\Delta^2_{\rm mf}(t)
\label{corrfnmf}
\end{eqnarray}
is also expected to show qualitatively similar behavior to $C(t)$.
Thus we conclude that the subsequent dynamics of the order parameter
following a quantum quench near a critical point is qualitatively
similar in nature to what is found from mean-field theory; however,
the precise quantitative value of, for example, its period of
oscillation, receives significant contribution from quantum
fluctuations.

\begin{figure}
\rotatebox{0}{\includegraphics*[width=0.85 \linewidth]{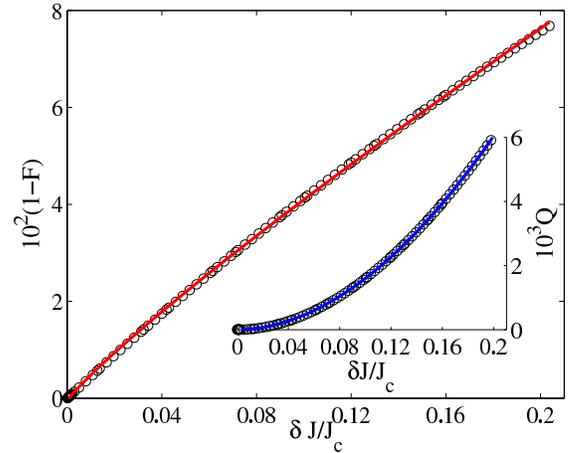}}
\caption{(Color online) Plot of $F$ and $Q$ as a function of the
$\delta J$ for $\delta J/J_c \ll 1$. The lines correspond to fits
yielding a power $1-F (Q) \sim (\delta J)^{r_1 (r_2)}$ with $r_1
\simeq 0.89$ and $r_2 \simeq 1.9$.} \label{fig3}
\end{figure}

Next, we compute the wavefunction overlap $F= |\langle
\psi_f|\psi_c\rangle|^2 = |\langle \psi'_f|e^{iS[J_f]}e^{-i S[J_c]}
|\psi'_c\rangle|^2$ for sudden quench starting at the QCP. Here
$\psi_f$($\psi_c$) denotes the ground state wavefunction for
$J=J_f(J_c)$. The residual energy $Q= \langle
\psi_c|\mathcal{H}[J_f]|\psi_c\rangle -E_G[J_f]$, where $E_G[J_f]$
denotes the ground state energy at $J=J_f$ as obtained by minimizing
$E$ in Eq.\ \ref{eneq1}, can also be computed in a similar manner.
Using the fact that for $|\psi'_c\rangle = e^{iS[J_c]}
|\psi_c\rangle$, $\varphi_{\bf r}=\Phi_{\bf r}=0$, we find, in terms
of the coefficients $f_n^{({\bf r})}$,
\begin{eqnarray}
Q &=&  E_G[J_c]-E_G[J_f] - 2 J \delta J \bar{n}(\bar{n}+1) \sum_{\bf
\langle {r r'}\rangle} \Big[
|f_{\bar{n}}^{({\bf r})}|^2 |f_{\bar{n}}^{({\bf r'})}|^2 \nonumber\\
&&- |f_{\bar{n}+1}^{({\bf r})}|^2 |f_{\bar{n}-1}^{({\bf r'})}|^2 -
f_{\bar{n}+1}^{*({\bf r})} f_{\bar{n}-1}^{({\bf r})}
f_{\bar{n}-1}^{*({\bf r'})} f_{\bar{n}+1}^{({\bf r'})}  \Big]/U.
\label{resen} \nonumber
\end{eqnarray}
A plot of $1-F$ and $Q$ for the homogeneous case, as a function of
$\delta J$ for $\delta J/J_c \lesssim 0.2$ is shown in Fig.\
\ref{fig3}. A numerical fit of these curves yields $1-F \sim \delta
J^{0.89}$ and $Q\sim \delta J^{1.90}$ which disagrees with the
universal scaling exponents ($1-F \sim \delta J^{d\nu}$ and $Q\sim
\delta J^{(d+z)\nu}$) expected from sudden dynamics across a QCP
with $z=1$ \cite{anatoly2}. In the next section, we shall study ramp
dynamics across the quantum critical point and try to understand the
reason behind such lack of universality in the dynamics of the
bosons.

\begin{figure}
\rotatebox{0}{\includegraphics*[width= 0.9 \linewidth]{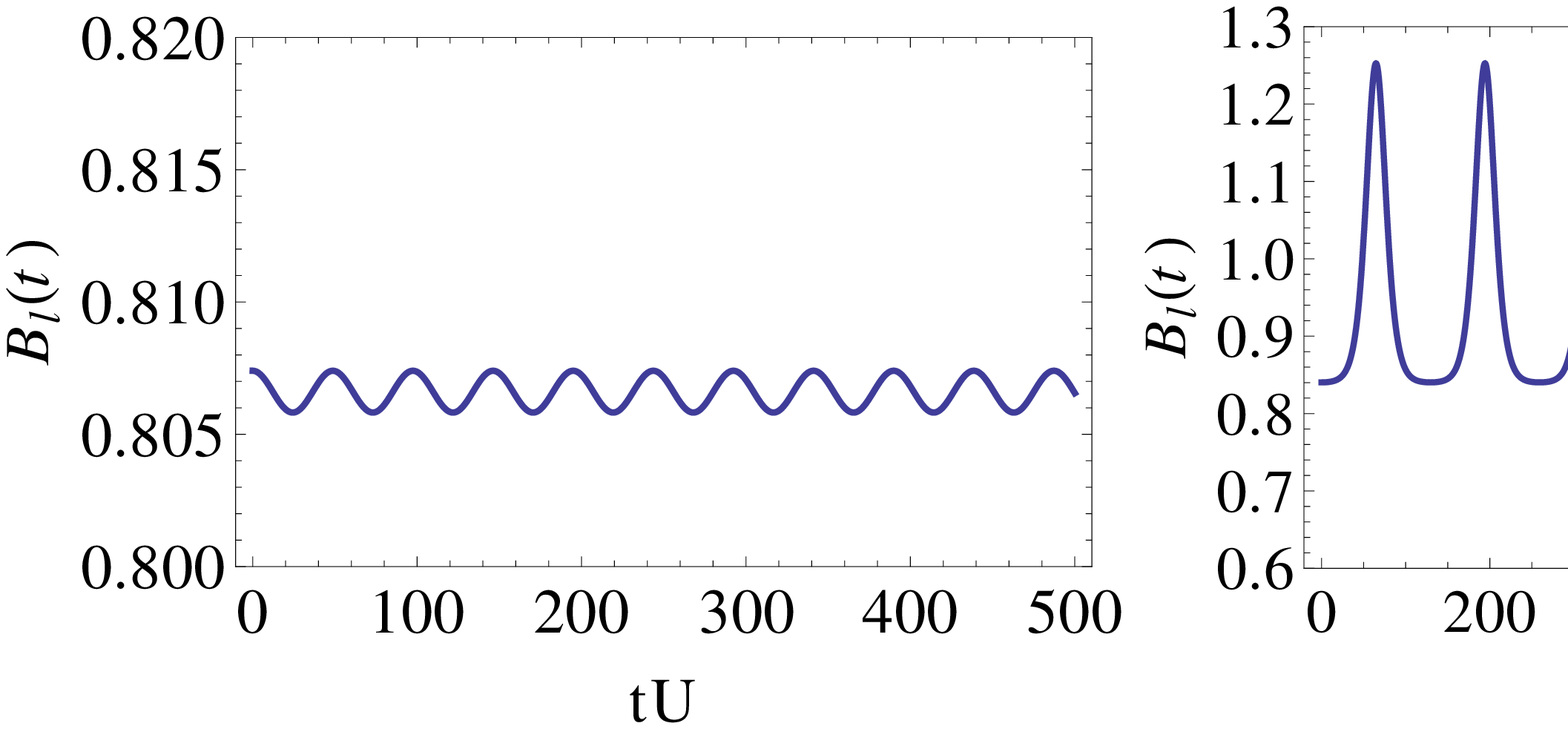}}
\caption{(Color online) Plot of $B_{\ell}(t)$ after a sudden quench
from $J_i$ to $J_f$ as a function of time for $J_f=0.98 J_c$ (left
panel) and $J_f=1.02 J_c$ (right panel). See text for details.}
\label{corrb}
\end{figure}

Before ending this section, we note that it is possible to compute
the evolution of the correlator  $B_{\ell} = \langle (b_{\bf
r}^{\dagger} b_{\bf r'} +{\rm h.c.})\rangle$ ($\ell$ is the link
between sites ${\bf r}$ and ${\bf r'}$) after a sudden quench from
$J=J_i$ to $J_f$ where $J_i$ corresponds to the MI state and $J_f$
correspond to either the SF phase or the MI phase. The mean-field
results for such a correlation would be zero if $J_f$ corresponds to
the MI state and $|\Delta^2(t)|$ if it corresponds to a homogeneous
SF phase. In contrast, using Eq.\ \ref{kecal} with $J=J_f$, we find
that the projection operator approach yields,
\begin{widetext}
\begin{eqnarray}
B_{\ell} &=& \Re{\left(\varphi_{\bf r}^*\varphi_{\bf
r'}\right)}+\frac{J_f}{U} \left[\Re{\left(\Phi_{{\bf
{r}},{\bar{n}-2}}^*\Phi_{{\bf r'},{\bar n}} +\Phi_{{\bf
r},{\bar{n}}} \Phi_{{\bf r'},{\bar{n}-2}}^*\right)}
+2\bar{n}\left(\bar{n}+1\right)|f_{\bar{n}}^{\left(\bf r\right)}|^2|f_{\bar{n}}^{\left(\bf r'\right)}|^2\right.\nonumber\\
&&\left.-\bar{n}\left(\bar{n}+1\right)\left(f_{\bar{n}-1}^{\left(\bf
r\right)} f_{\bar{n}+1}^{{\left(\bf
r\right)}*}f_{\bar{n}+1}^{{\left(\bf
{r'}\right)}}f_{\bar{n}-1}^{{\left(\bf {r'}\right)}*}
+f_{\bar{n}+1}^{\left(\bf r\right)} f_{\bar{n}-1}^{{\left(\bf
r\right)}*}f_{\bar{n}-1}^{{\left(\bf
{r'}\right)}}f_{\bar{n}+1}^{{\left(\bf {r'}\right)}*}+
|f_{\bar{n}+1}^{\left(\bf r\right)}|^2|f_{\bar{n}-1}^{\left(\bf {r'}\right)}|^2\right.\right.\nonumber\\
&&\left.\left.+|f_{\bar{n}-1}^{\left(\bf
r\right)}|^2|f_{\bar{n}+1}^{\left(\bf {r'}\right)}|^2\right)\right]
+ \frac{J_f}{U} \sum_{\langle {\bf r} {\bf r'} {\bf r''} \rangle}
\Big \{ 2\Re \Big[ \varphi_{{\bf r},\bar{n}-1}^*(\bar{n}+1)
|f_{\bar{n}}^{({\bf r'})}|^2\nonumber \\
&& + \varphi_{{\bf r}\bar{n}} \Phi_{{\bf r'},\bar{n}-2}^* -
\varphi_{{\bf r}\bar{n}}^* \bar{n} |f_{\bar{n}-1}^{({\bf r'})}|^2 -
\varphi_{{\bf r},\bar{n}-1} \Phi_{{\bf r'},\bar{n}-1}^* \Big]
\varphi_{{\bf r''}} + 2\Re \Big[ \varphi_{{\bf r},\bar{n}-1}^*
\Phi_{{\bf r'}\bar{n}} + \varphi_{{\bf r}\bar{n}}
\bar{n}|f_{\bar{n}}^{({\bf r'})}|^2\nonumber \\
&&- \varphi_{{\bf r}\bar{n}}^* \Phi_{{\bf r'},\bar{n}-1} -
\varphi_{{\bf r},\bar{n}-1}(\bar{n}+1) |f_{\bar{n}+1}^{({\bf
r'})}|^2  \Big] \varphi_{{\bf r''}}^*\Big \}, \label{corela}
\end{eqnarray}
\end{widetext}
When $J_f$ correspond to the MI phase, since $f_{{\bar n}}(t) \gg
f_{n\ne {\bar n}}(t)$, we find that $B_{\ell} $ shows very small
oscillations around the base value $4 J_f |f_{\bar n}|^4/U$. In
contrast, it displays significant oscillation in the SF phase. These
behaviors, in the homogeneous limit, are sketched in the left and
right panels of Fig.\ \ref{corrb} for $J_f =0.98 J_c$ (left panel)
and $1.02 J_c$ (right panel). We note that the behavior of
$B_{\ell}$ in the MI phase is qualitatively different from the
mean-field result which correspond to the first term in the right
side of Eq.\ \ref{corela}.

\subsection{Non-linear Ramp}

In this section, we address the dynamics of the bosons during a ramp
of the hopping amplitude $J$ characterized by a rate $\tau^{-1}$ and
an exponent $\alpha$: $J(t)=J_i +(J_f-J_i) (t/\tau)^{\alpha}$. Note
that the system evolves from $J_i$ at $t_i=0$ to $J_f$ at $t_f
=\tau$; consequently as long as we restrict ourselves to $J_i/U,
J_f/U \ll 1$, we expect the perturbative projection method to
address the dynamics accurately irrespective of the values of $\tau$
and $\alpha$. Thus the projection operator method enables one to
address ``slow" and ``fast" and linear/non-linear ramps at equal
footing. We note at the outset that our results in this section
reproduce those in Ref.\ \onlinecite{ct1} as a special case for
$\alpha=1$.

\begin{figure}
\rotatebox{0}{\includegraphics*[width= 0.9 \linewidth]{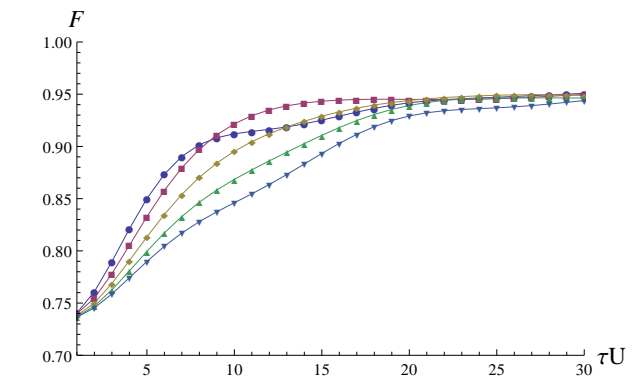}}
\caption{(Color online) Plot of $F=1-P$ as a function $\tau U$ (in
units of $\hbar=1$) for $J_i/U=0.05$ (SF phase) and $J_f/U=0.005$
(Mott phase) for $\alpha=1$ (blue circles), $2$ (red squares), $3$
(yellow diamonds), $4$ (green triangles), and $5$ (blue inverted
triangles) showing the plateau-like behavior at large $\tau$. }
\label{fig4}
\end{figure}

\begin{figure}
\rotatebox{0}{\includegraphics*[width= 0.9 \linewidth]{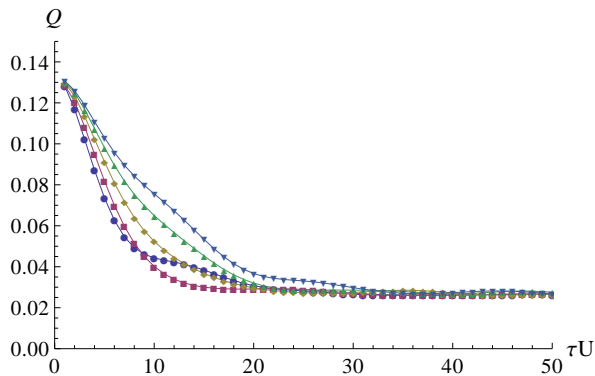}}
\caption{(Color online) Plot of $Q$ as a function of $\tau$ for
$\alpha=1 .. 5$. All parameters and symbols are same as those in
Fig.\ \ref{fig4}} \label{fig5}
\end{figure}

\begin{figure}
\rotatebox{0}{\includegraphics*[width= 0.9 \linewidth]{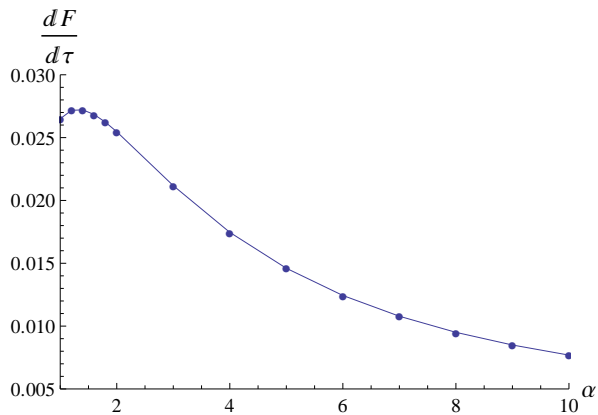}}
\caption{(Color online)  Plot of $dF/d\tau$ for small $\tau$ ($\tau
U \le 5$) as a function of $\alpha$.} \label{fig6}
\end{figure}

To address the dynamics, we use Eq.\ \ref{feq1} and solve for
$f_n^{({\bf r})}\equiv f_n$ for translationally invariant systems.
This enables us to compute the defect formation probability $P=
1-|\langle \psi_G|\psi(t_f)\rangle|^2=1-F$, where $|\psi_G\rangle$
($|\psi(t_f)\rangle$) denotes the final ground state (state after
the ramp), for a ramp from $J_i/U=0.05$ (superfluid phase) to
$J_f/U=0.005$ (Mott phase) as a function of $\tau$. The behavior of
$F$ and $Q$ are shown in Figs.\ \ref{fig4} and \ref{fig5} for
various representative values of $\alpha$. We find that both $Q$ and
$F$ (and hence $P$) exhibits a plateau like behavior at large
$\tau$. The slope of both $F$ and $Q$ for small $\tau$ depend on the
ramp protocol through the exponent $\alpha$; however, the asymptotic
values of these quantities at large $\tau$ is independent of
$\alpha$. The plot of $dF/d\tau$ for $\tau U \le 5$ (where $F$ is
approximately linear in $\tau$ as can be seen from Fig.\ \ref{fig4})
is shown in Fig.\ \ref{fig6}. The slope decreases monotonically with
$\alpha$ for large $\alpha$ which indicates that $F$ (and similarly
$Q$) saturates at larger values of $\tau$ with increasing $\alpha$.
The slope is maximal, indicating minimal initial defect production,
for $\alpha=1.5$.

It is clear from the plots that both $P$ and $Q$ do not display
universal scaling as expected from generic theories of slow dynamics
of quantum systems near critical point \cite{rev1}. This seems to be
in qualitative agreement with the recent experiments presented in
Ref.\ \onlinecite{exp1}, where linear ramp dynamics of ultracold
bosons from superfluid to the Mott region has been experimentally
studied. Indeed, it was found, via direct measurement of parity of
${\bar n}$ per site, that $F$ displays a plateau like behavior
similar to Fig.\ \ref{fig4}. Such a lack of universality in the
dynamics can be qualitatively understood from absence of
contribution of the critical (${\bf k}=0$) modes. In the
strong-coupling regime ($z_0J/U \ll 1$), the system can access the
${\bf k}=0$ modes after time ${\mathcal T}$ which can be roughly
estimated as the time taken by a boson to cover the linear system
dimension $L$. For typical small $J$ ($U=1$) in the Mott phase and
near the QCP, ${\mathcal T} \sim {\rm O}(L\hbar/J)$ can be very
large. Thus for $t\le{\mathcal T}$, the dynamics, governed by local
physics which is well captured by our method, do not display
critical scaling behavior. We note that our theory which is based on
building on spatial correlation order by order in powers of $z_0J/U$
shall not easily capture the physics associated with long-range
spatial correlation near the critical point and will deviate from
experimental results for much slower ramp rates. It seems, however,
that achieving such low ramp rates for the present system in the
Mott phase can be experimentally challenging.

\section{Discussion}

\label{dissec}

In conclusion, we have presented a projection operator formalism
that describes in a semi-analytical way both the phase diagram and
non-equilibrium dynamics of the Bose-Hubbard model. It produces a
phase diagram which is nearly identical to the QMC results in 3D,
allows for a computation of the low-energy excitation spectra of the
system, and yields semi-analytical insight for several quantities
such as $F$, $Q$, $\Delta(t)$, $P$, and $C(t)$ for non-equilibrium
dynamics. Its prediction for $P$ for a slow ramp matches
qualitatively with recent experiments. The method, in principle, can
be generalized to any strongly correlated systems which allows
perturbative treatment of fluctuations. We leave such considerations
for future study. We also note that studying finite temperature
physics of the Bosons with our method also poses an interesting
theoretical challenge. For now, we can only estimate the range of
physical temperatures $T$ for which the $T=0$ theory is accurate.
For a typical lattice depth in the Mott or critical regime, one can
estimate $U \sim 2$ kHz $\simeq 200$nK \cite{bloch1}. This yields,
in 3D, a melting temperature $T^{\ast} \simeq 0.2 U = 40$nK for the
Mott phase and critical temperature $T_c \simeq z_0J_c \simeq 35$nK
for the SF phase at Mott tip \cite{ketterle01}. This necessitates $T
\ll T_c, T^{\ast}$ to be a few nano-Kelvins which is well within the
current experimental limit $\sim 1$nK \cite{ketterle01}.

CT thanks M. Lewenstein for support during the work. The authors
thank E. Altman, C. Lannert, S. Mondal, A. Polkovnikov, R. Sensarma,
and S. Vishveshwara for discussions on related topics, and B.
Caprogrosso-Sansone for sharing QMC data. KS thanks DST, India for
support under Project No. SR/S2/CMP-001/2009. CT acknowledges
support of Spanish MEC (FIS2008-00784, QOIT) and hospitality of
Theoretical Physics Department, IACS.

\appendix
\section{Explicit form of the fields}
\label{app:fields} In this section we provide explicit expressions
for the fields used in Eq~\ref{feq1}. In what follows we define
$\delta_{ij}$ to be the Kronecker delta and
$\Lambda_{ij}=1-\delta_{ij}$. The fields that multiply the time
derivative of the tunneling coefficient are given by
\begin{eqnarray}
\overline{\partial\alpha}_i &=& \sum_{\langle b \rangle_i} \Big( \delta_{n\bar{n}} \varphi_{b\bar{n}} - \delta_{n,\bar{n}+1} \varphi_{b,\bar{n}-1}\Big),\\
{\overline{\partial\beta}_i}^* &=& \sum_{\langle b \rangle_i} \Big( \delta_{n\bar{n}} \varphi_{b,\bar{n}-1}^* - \delta_{n,\bar{n}-1} \varphi_{b\bar{n}}^*  \Big).
\end{eqnarray}
The phase factor $\chi_{\rm n}^{(i)}$ is given by
\begin{multline}
\chi_{\rm n}^{(i)} = -\mu n + \frac{U}{2}n(n-1) - \frac{2J^2}{U} \delta_{n\bar{n}} \bar{n}(\bar{n}+1) \sum_{\langle a \rangle_i} |f_{\rm \bar{n}}^{(a)}|^2 \\
- \frac{2J^2}{U} \sum_{\langle a \rangle_i}\sum_{\langle c \rangle_i}\Lambda_{ac}\Big[ \bar{n} (\delta_{n\bar{n}} - \delta_{n,\bar{n}-1}) \Re{\varphi_{a\bar{n}}\Big(\varphi_{c}^* - \frac{\varphi_{c\bar{n}}^*}{2}\Big) }  \\
+ (\bar{n}+1) (\delta_{n\bar{n}} - \delta_{n,\bar{n}+1}) \Re{\varphi_{a,\bar{n}-1}}\Big(\varphi_{c}^* - \frac{\varphi_{c,\bar{n}-1}^*}{2} \Big) \Big]  \\
+ \frac{J^2}{U}\bar{n}(\bar{n}+1)\Big[ \delta_{n,\bar{n}+1} \sum_{\langle a \rangle_i}|f_{\rm \bar{n}-1}^{(a)}|^2 + \delta_{n,\bar{n}-1} \sum_{\langle a \rangle_i}|f_{\rm \bar{n}+1}^{(a)}|^2\Big].
\end{multline}
Notice that $\chi_{\rm n}^{(i)}$ gives only a phase factor in real
time dynamics and is therefore negligible; however, it is important
in the imaginary time evolution. The fields $\overline{\alpha}_\rr$
and ${\overline{\beta}_\rr}^*$, which couples $\partial_t f_{\rm
n}^{(\rr)}$ to $f_{\rm n\pm1}^{(\rr)}$ linearly in $J$ are given by:
\begin{multline}
\overline{\alpha}_i = \sum_{\langle a \rangle_i} \Big[ \varphi_{a} + \frac{J}{U} \mathcal{A}_a^{(i)} -\delta_{n\bar{n}} \Big(\varphi_{a\bar{n}} - \frac{J}{U}\mathcal{B}_a^{(i)} \Big) \\
-\delta_{n,\bar{n}+1} \Big(\varphi_{a,\bar{n}-1} - \frac{J}{U}\mathcal{C}_a^{(i)}\Big) \Big],
\end{multline}
and
\begin{multline}
{\overline{\beta}_i}^* = \sum_{\langle a \rangle_i} \Big[ \varphi_{a}^* + \frac{J}{U}\mathcal{A}_a^{*(i)} -\delta_{n,\bar{n}-1} \Big(\varphi_{a\bar{n}}^* - \frac{J}{U}\mathcal{B}_a^{*(i)} \Big) \\
-\delta_{n\bar{n}} \Big(\varphi_{a,\bar{n}-1}^* - \frac{J}{U}\mathcal{C}_a^{*(i)}\Big)\Big].
\end{multline}
In contrast, the fields $\overline{\eta}_\rr$ and
${\overline{\xi}_\rr}^*$, which couple $\partial_t f_{\rm
n}^{(\rr)}$ to $f_{\rm n\pm2}^{(\rr)}$ to ${\rm O}(J^2/U)$ are given
by
\begin{multline}
\overline{\eta}_i = \sum_{\langle a \rangle_i} \Big\{ \delta_{n\bar{n}}\Phi_{a\bar{n}} + \delta_{n,\bar{n}+2}\Phi_{a,\bar{n}-2} - \delta_{n,\bar{n}+1} \Phi_{a,\bar{n}-1} \\
+\sum_{\langle c \rangle_i}\Lambda_{ac} \Big(\delta_{n\bar{n}}\varphi_{a\bar{n}} + \delta_{n,\bar{n}+2}\varphi_{a,\bar{n}-1} \Big) \varphi_{c} \nonumber \\
- \delta_{n,\bar{n}+1}\sum_{\langle c \rangle_i}\Lambda_{ac} \Big[ \Big( \varphi_{a\bar{n}} + \varphi_{a,\bar{n}-1}\Big) \varphi_{c} - \varphi_{a\bar{n}}\varphi_{c,\bar{n}-1} \Big] \Big\},
\end{multline}
and
\begin{multline}
{\overline{\xi}_i}^* = \sum_{\langle a \rangle_i} \Big\{\delta_{n,\bar{n}-2} \Phi_{a\bar{n}}^* + \delta_{n\bar{n}} \Phi_{a,\bar{n}-2}^* - \delta_{n,\bar{n}-1} \Phi_{a,\bar{n}-1}^*\\
+ \sum_{\langle c \rangle_i}\Lambda_{ac} \Big( \delta_{n,\bar{n}-2}\varphi_{a\bar{n}}^* + \delta_{n\bar{n}}\varphi_{a,\bar{n}-1}^*\Big) \varphi_{c}^*  \\
- \delta_{n,\bar{n}-1}\sum_{\langle c \rangle_i}\Lambda_{ac} \Big[ \Big( \varphi_{a\bar{n}}^* + \varphi_{a,\bar{n}-1}^*\Big) \varphi_{c}^* - \varphi_{a\bar{n}}^*\varphi_{c,\bar{n}-1}^* \Big]\Big\},
\end{multline}
where we have introduced the quantities
\begin{multline}
\mathcal{A}_a^{(i)} = \sum_{\langle b \rangle_a}\Lambda_{bi} \Big[
\Big( \Phi_{a,\bar{n}-2} - \Phi_{a,\bar{n}-1}\Big) \varphi_{b\bar{n}}^* + \Big( \Phi_{a\bar{n}} - \Phi_{a,\bar{n}-1}\Big) \varphi_{b,\bar{n}-1}^* \\
+ (\bar{n}+1) \Big( |f_{\rm \bar{n}}^{(a)}|^2 - |f_{\rm \bar{n}+1}^{(a)}|^2\Big)\varphi_{b,\bar{n}-1}
+ \bar{n} \Big( |f_{\rm \bar{n}}^{(a)}|^2 - |f_{\rm \bar{n}-1}^{(a)}|^2\Big) \varphi_{b\bar{n}} \Big],
\end{multline}
\begin{multline}
\mathcal{B}_a^{(i)} = \sum_{\langle b \rangle_a}\Lambda_{bi} \Big[
(\bar{n}+1) \Big( |f_{\rm \bar{n}}^{(a)}|^2 - |f_{\rm \bar{n}+1}^{(a)}|^2\Big) \big( \varphi_b - \varphi_{b,\bar{n}-1} \big)\\
+ \Big( \Phi_{a\bar{n}} - \Phi_{a,\bar{n}-1}\Big) \varphi_{b}^* +  \Phi_{a,\bar{n}-1} \varphi_{b\bar{n}}^* \Big],
\end{multline}
and
\begin{multline}
\mathcal{C}_a^{(i)} = \sum_{\langle b \rangle_a}\Lambda_{bi} \Big[
\bar{n} \Big( |f_{\rm \bar{n}}^{(a)}|^2 - |f_{\rm \bar{n}-1}^{(a)}|^2\Big) \big( \varphi_{b} - \varphi_{b\bar{n}}\big)\\
+ \Big( \Phi_{a,\bar{n}-2} - \Phi_{a,\bar{n}-1}\Big) \varphi_{b}^*+
\Phi_{a,\bar{n}-1} \varphi_{b,\bar{n}-1}^* \Big],
\end{multline}
for notational convenience.

\end{document}